# FUTURE PHOTOVOLTAIC ELECTRICITY PRODUCTION TARGETS AND THE LINK TO CONSUMPTION PER CAPITA ON THE POLICY LEVEL IN MENA REGION

Mostafa Abdelrashied, Dikshita Bhattacharya, Pouzant Baliozian, Ralf Preu
Albert-Ludwig University of Freiburg

**ABSTRACT:** This paper provides an overview of the status of the electricity market in the region, indicating the nexus between electricity consumption with population growth and GDP. It also analyzes the policy portfolio in different countries, indicating some of the in-action policies' effectiveness and recommended alternatives. World Bank datasets were used for the analysis between 2000 and 2014. We found that the MENA region is at an early stage for renewable energy with a high potential for solar energy, making it attractive for investors. However, the high dependency on oil for consumption and exporting might not provide a prosperous environment for renewable technologies to grow. Therefore, a greater focus on decoupling economic growth from energy consumption will have a long-lasting impact on fiscal revenues for net-oil exporting countries. Moreover, the consequences of the decoupling will allow more renewables penetration in the current energy mix enabling many countries to reach their Paris Agreement goals. For short-term energy policy actions, starting a subsidy reform towards the final repeal of subsidies is a must as these measures relate to all end-use sectors and impact fiscal stability in many countries. With its 1.65GW Benban Solar Park in Aswan, Egypt has shown an example of shifting from subsidizing fossil fuel products to commissioning renewable projects to get closer to its Paris Agreement targets.

## 1. INTRODUCTION

The MENA countries have about 70% of the world's proven oil reserves and 40% of the proven gas reserves [1], which is termed as the black gold as it was and is a key contributor to the transformation of a lot of these countries from isolated deserts to the urbanization. Some countries depend on fossil-fuel revenues for up to 90% of their GDP [2]. The recent fluctuation of oil prices reduced these revenues, creating pressure to reduce state spending, causing macroeconomic volatility and massive deficiency in the governmental budget [3]. Moreover, it revealed a high degree of abnormal and unreasonable inefficiency levels of energy consumption. Seemingly, they realized that a very high consumption with inequivalent increase in GDP will bring an end to this prosperity. Therefore, different countries have initiated plans to shift their dependency from fossil-fuel resources to renewable energy, mainly solar energy. With the current heavily subsided economy, it is difficult for renewables to compete or have a high share in the energy mix.

The irradiation in the MENA is one of the highest in the whole world. It ranges between 2000 – 2500 kWh/m2, nearly double the solar radiation of western-EU countries, e.g., Germany [5]. Moreover, the lowest recorded cost for utility-scale solar PV was announced to provide electricity with $0.0584 per kWh in Dubai [6]. The REACCESS project 'Risk of Energy Availability: Common Corridors for Europe Supply Security' illustrated that 15% to 20% of the total EU-27 electricity demand by 2050 could be covered by imports of renewable electricity from countries of the MENA region [7]. Hence, a comprehensive policy structure and legal framework are essential for a secure, environmentally friendly, and economically efficient energy supply system.

This paper investigates the nexus between electricity consumption, GDP and population growth for the MENA countries. Consequently, the impact of the recent energy policy changes on the PV future in the region due to the expected increase in electricity consumption, indicating the instrumental policies in use for PV and shedding light on what is missing. Furthermore, providing data with the concurrent PV share into regions' electricity markets and the announced targets to be achieved in 2020 and 2030 plans.

## 2. METHODOLOGY.

A quantitative data collection approach was used to investigate the impacts of politics on the renewable energy sector, particularly on PV (Figure 1). Analysis of the available data using this approach will demonstrate in-depth insight into how the governments are advancing to renewables, how efficient they are, and the parameters that govern this behavior. Only datasets available on the world bank and International Energy Agency (IEA) have been used to avoid bias in governmental reports. A contemporary data gathering has been tracked to avert misleading explanations for the projections in the future. Therefore, only papers and articles from 2014 and later had been reviewed and analyzed; otherwise, it will be used only as a preview for explaining historical behaviors.

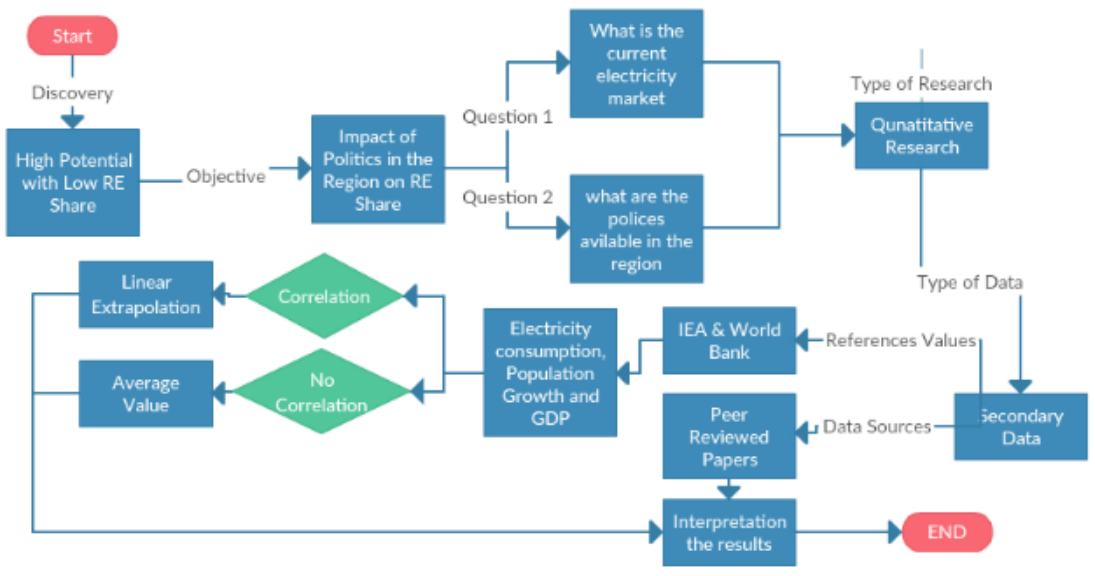

**Figure 1.** Flow chart of the used methodology

## 3. DEFINING THE MENA REGION

The MENA region has no standard definition to be followed, it depends on the organization that studies the region and what is the topic of interest. For this paper, the definition given by the World Bank [8] will be used (Figure 2) due to the abundance of data for those countries.

- North Africa (or Maghreb): Morocco, Algeria, Tunisia, Libya, and Egypt
- Mediterranean (or Levant): Jordan, Israel, Lebanon, and Syria
- Gulf Cooperation Council: Bahrain, Kuwait, Oman, Saudi Arabia, Qatar, and United Arab Emirates

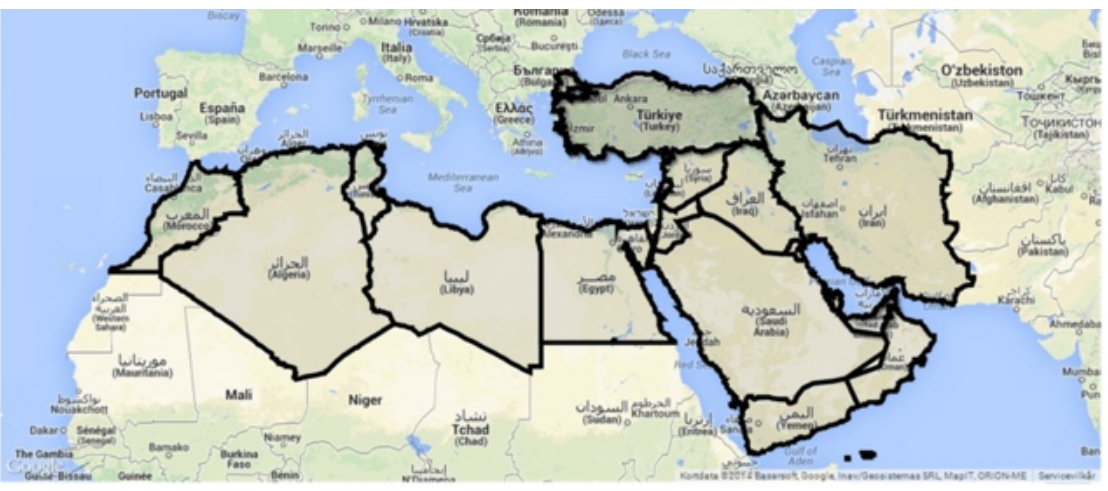

- Another Middle East countries: Iran, Iraq, Djibouti, and Yemen

**Figure 2**. Countries of the MENA region (taken from Ref [https://maps.google.com/])

## 4. PV POTENTIALS ASSESSMENT IN THE MENA

It is necessary to evaluate countries' actual performance according to their potential limit. Though, assessing the potentials for PV is tricky for several reasons. Solar irradiation is not evenly distributed in all the countries, the available area for PV utilities varies from one country to another, and the electricity prices differ considerably between regions according to the political & economic situation. Therefore, a unified methodology is needed for fair and objective comparison between different countries in the region. Continuing with the World Bank data collection approach, the same methodology for PV potential assessment globally will be used for the MENA region [9].

### 4.1. THEORETICAL POTENTIAL

It is defined as the total primary solar energy flux hitting the earth's surface suitable for PV. In contrast to CSP, PV uses both direct and diffuse components of solar irradiation. Global Horizontal Irradiation (GHI) will be used to evaluate the maximum solar energy reaching different countries. Countries like Saudi Arabia, Oman, Yemen, large parts of Egypt, Algeria, Libya, and Morocco have shown very high GHI rates, which make them ideal for PV generation in terms of solar irradiation.

### 4.2. GEOGRAPHICAL POTENTIAL

The geographical potential is defined as the land area available for PV installation. The region has large deserts with very high temperatures, leading to arid conditions for agriculture and challenging conditions for urban development. Despite these harsh conditions, modified solar panels have been developed to withstand the dust and humidity of the region without degradation of efficiency due to the extreme temperatures. For PV projects, it is not very easy to say if deserts are suitable for PV power plants or not. The significant advantages of setting up a PV project in a desert include the large available areas, high average solar radiation, high energy per capita consumption, and no shading. However, extreme temperatures, Logistics, grid connectivity problems, additional investment for cooling, and costly maintenance are some of the major setbacks to be addressed.

### 4.3. TECHNICAL POTENTIAL

Technical potentials can be measure as the geographic potential after any efficiency losses from the primary to secondary conversion processes. It is measured by the PV electricity output (PVout), which is the amount of energy converted by a PV system into electricity [kWh/kWp] that is expected to be generated according to the geographical conditions and the PV system configuration. These values are very high in the MENA region ranging between 1900kWh/kWp to 2400 kWh/kWp [7].

### 4.4. ECONOMICAL POTENTIAL

It is the portion of the technical potential that is competitive with another relevant form of electricity generation. Hence, the cost of PV electricity cannot be isolated from its competitiveness. Therefore, the annual Levelized Electricity Cost (LCE) production of a PV System is calculated and used for evaluation [10, 11]. These prices used to be very high compared to other conventional power plants due to the subsidies for these fuels, but after the recent reforms in the region, they reach low values; these values are among the lowest in the world since the lowest recorded cost for utility-scale solar PV was announced to provide electricity reaching $0.0584 per kWh in Dubai [6].

## 5. ELECTRICITY MARKET OVERVIEW:

The MENA region has minimal coal reserves but abundant oil and gas reserves, causing the non-renewable resources to dominate electricity generation. Saudi Arabia was the highest contributor, with 340 TWh in 2015 [12] due to high energy-intensive industries and high consumption cooling loads. In Iran and Egypt, their trend is similar, mainly due to the high population growth rate reaching almost 100M in Egypt [13] and 82M in Iran [14]. When looking at the energy consumption per capita, the countries' ranks will be different, which indicates a huge gap in energy efficiency (Figure 3). Gulf countries have the highest levels. The high subsidies on fossil fuels consumption cause the end prices to be fractions of the international prices. Hence, a very high degree of inefficiency will be the primary trend for the consumers.

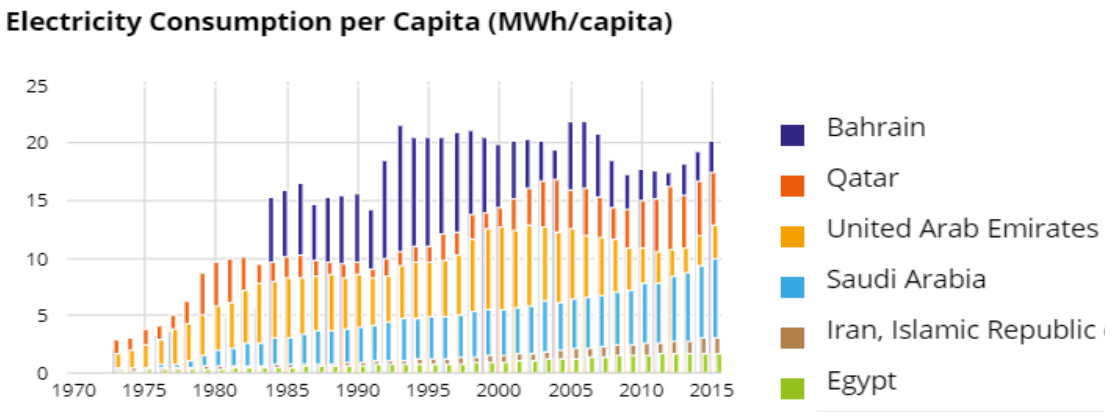

**Figure 3**. Electricity consumption per capita [13]

The electricity production from non-renewable excluding hydropower for the whole region is merely reaching 0.5% of the total electricity production in 2014. The highest contributor for this share is Morocco with 6.7%, and Tunisia is far behind with 2.8%. [15]. This meager contribution for renewables is caused by many barriers, mainly due to the energy market structures. These barriers can be classified as follows [10, 16–18].

- Market Barriers
  - Lack of broad information/awareness by decision-makers
  - Subsidies for conventional energies create market distortion
- Regulatory/administrative barriers
  - Lack of adequate institutional and regulatory framework
  - Absence of systematic policy support
- Financial barriers
  - High up-front costs for investors – especially for off-grid applications compared to the conventional technologies
  - High cost of capital for RES investments
- Other barriers
  - The lack of adequate transmission and distribution grids needed to properly integrate the PV systems into the electrical grid.

The projected increase for capacity anticipated for different countries in the region varies according to the population's GDP growth rate. Saudi Arabia and Iran installed capacity will increase, reaching 100 GW and 90 GW respectively in 2020 due to the expected growth rate and the expected industrial activities. However, the huge leap - nearly the double - of installed capacity for Egypt will occur to meet the demand generated from the high population growth and the expected high GDP growth rate in the future after the recent great depression [6]

## 6. RESULTS – GDP, POPULATION GROWTH AND ELECTRICITY CONSUMPTION NEXUS

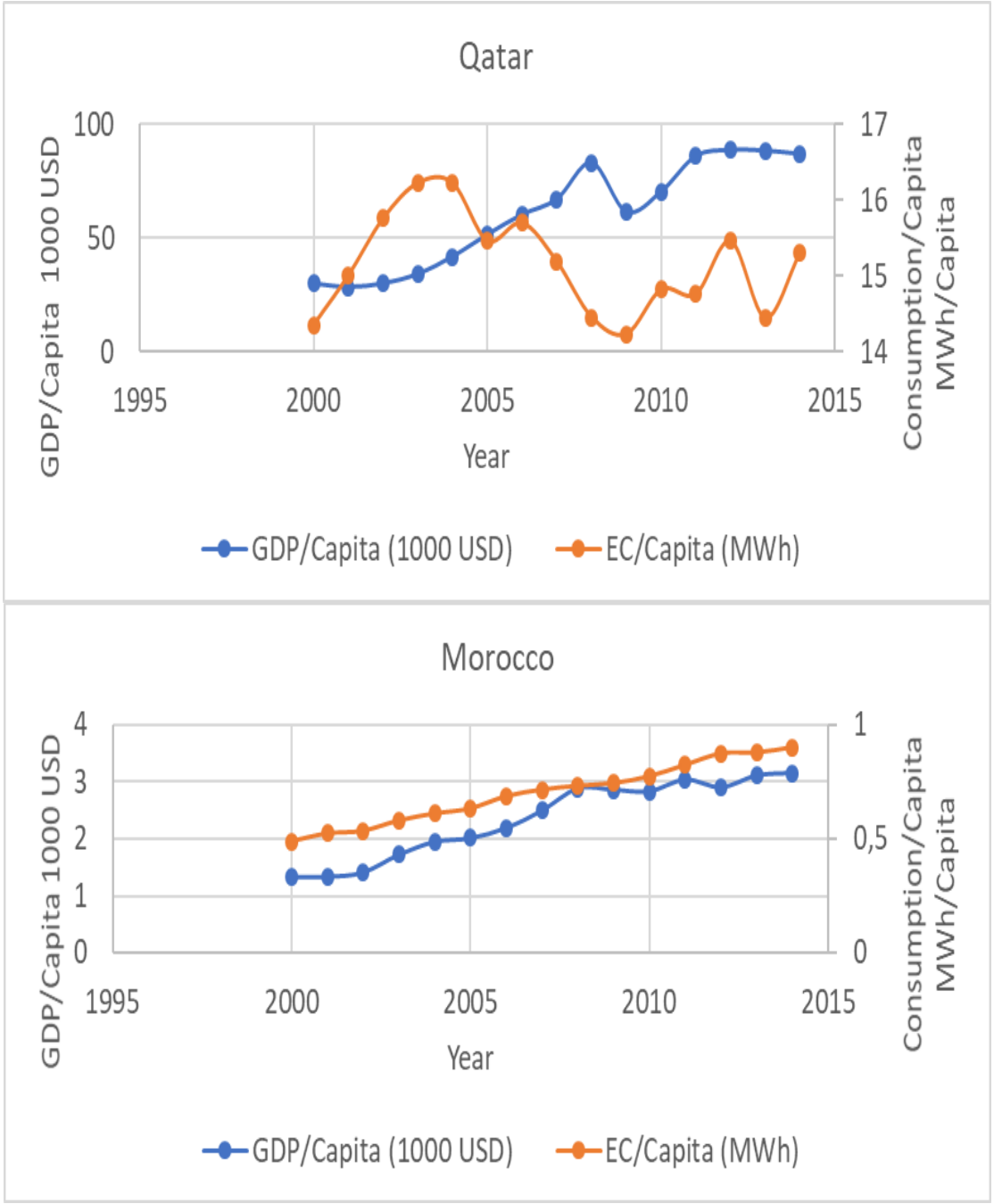

*Figure 4-5. Correlation between GDP and EC per capita*

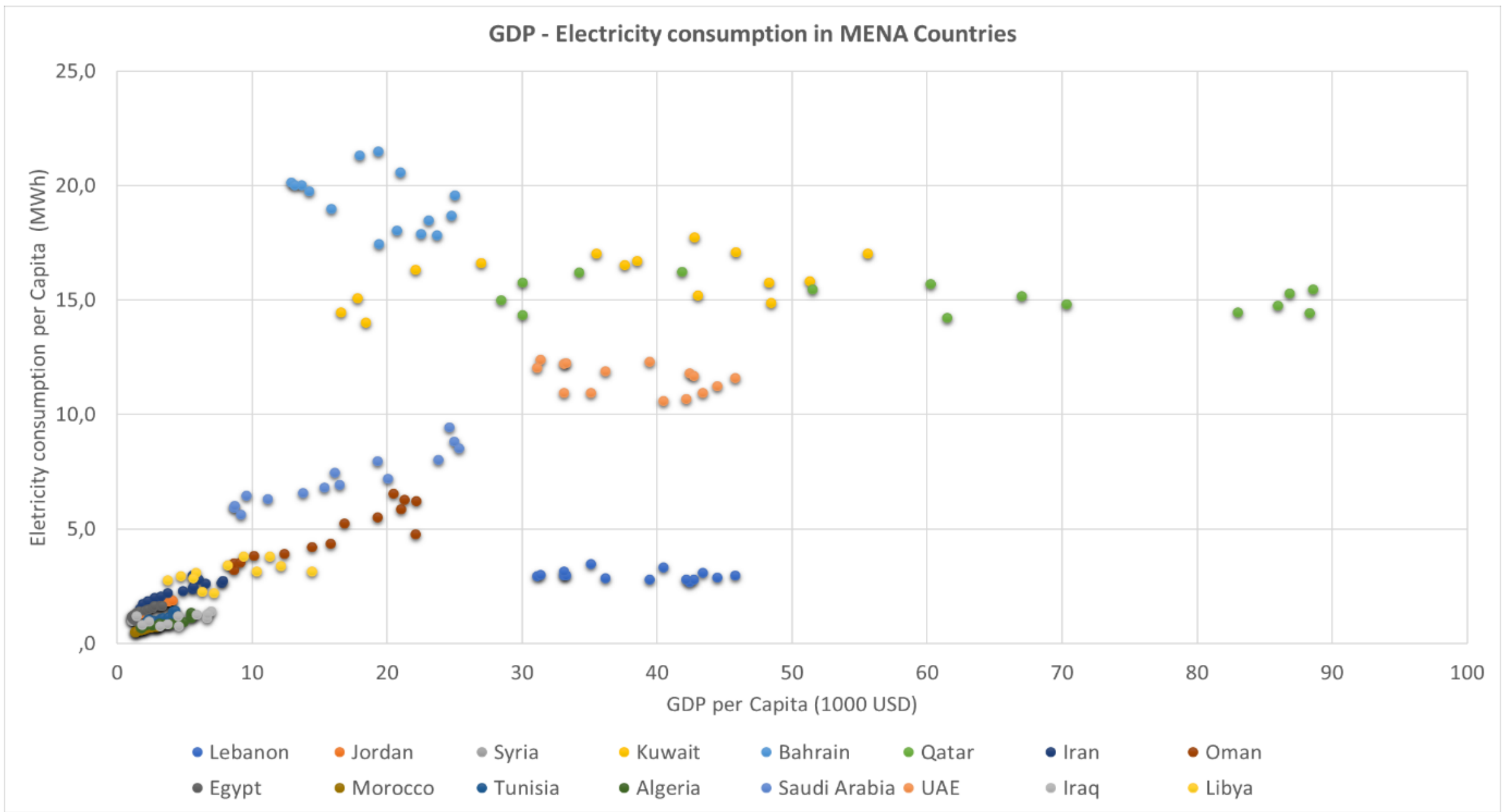

*Figure 6. MENA countries GDP correlation with electricity consumption*

This section investigates the relationship between economic growth, population, and electricity consumption between 2000 to 2014. many studies found a unidirectional causality between electricity consumption and economic growth, while some studies found that there was bidirectional causality [19–23]. Figure 4 and 5 show both directional causalities exist depending on the country. The GCC countries tend to the bidirectional causality, but North Africa and Levant Countries tend to have unidirectional causality. For instance, Qatar showing a positive correlation with a dip during 2008 crisis between GDP growth and Electricity Consumption is an example of GCC behavior. Morocco, on the other hand, shows a linear relationship between GDP and electricity consumption. For a comprehensive relationship between the GDP and electricity consumption in these countries, a curve correlating between GDP & Electricity consumption is plotted for each country between 2000 and 2014 (Figure 5). The GCC behaves similarly in a particular trend, and the North Africa region is behaving nearly like the Levant countries except Lebanon, with a trend of having low GDP and electricity consumption per capita compared to the GCC Region. The GCC countries benefit from the wealth of oil-exporting to get a high GDP per capita, reaching nearly 85,000 USD. Besides this high GDP per capita, the electricity consumption also tends to increase even compared with countries with the same GDP, for example, Lebanon and Kuwait. Lebanon can be considered a noticeable outlier as it shows middle GDP per capita with extremely low electric consumption. This can be explained in the light of the small-scale diesel electric generators people use to combat frequent blackouts and shortages.

## 7. RESULTS – ENERGY POLICY PORTFOLIO IN THE REGION

Considering these countries' social, political, and economic circumstances is essential as these elements impact their energy systems. The net-oil exporting countries are primarily reliant on hydrocarbon exports for overall economic activity and fiscal revenue. The result of such natural resource dependence is Macroeconomic volatility, Lack of incentive for establishing regulatory and institutional frameworks supportive of solid private sectors. Financing abilities will differ considerably between countries depending on their oil reserves. International commercial banks will be more than willing to finance renewable energy projects in oil-rich countries due to their high credit rating and the high quality of the project sponsors and off-takers. However, the credit ratings of countries without oil or natural gas reserves are significantly worse, meaning that participation from multilateral financial organizations and export credit agencies will be critical to unlocking commercial debt. A summary of the available renewable energy policies with the following targets in 2020 and 2030 is evaluated and tabulated (Table 3). These targets are divided into two visions in most countries, vision 2020 and vision 2030. As shown in (Figure 5), the most ambitious country in the region is Morocco with a 60% percent target of renewables, in the second place in Egypt to meet the expected high demand from the high population.

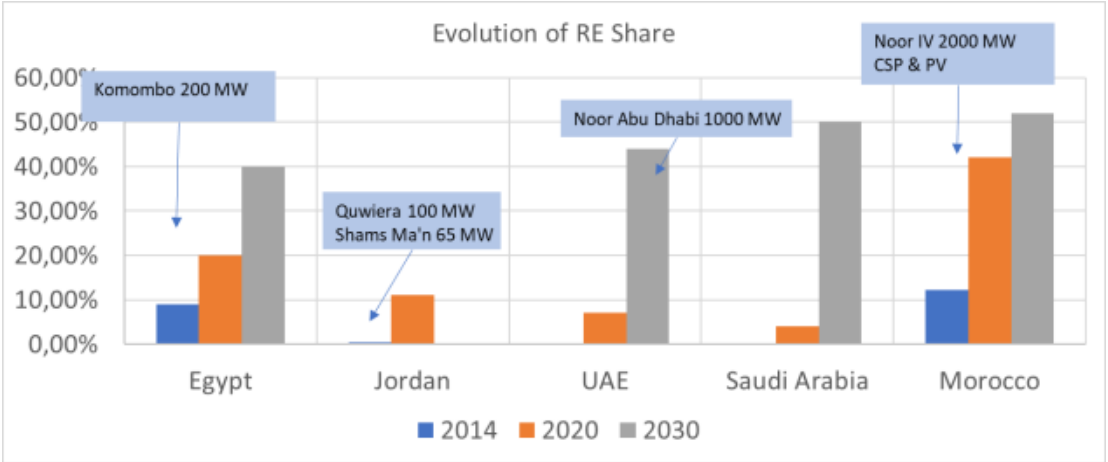

**Figure 5.** Targets for RE in 2020 and 2030

## 8. RESULTS – ENERGY POLICY REFORM

During the last three years, decisions had been made by some governments regarding energy prices. Starting from Egypt, after the Arab Spring, all successive governments avoided any decisions that may lead to protests, not only this but also reviving the endeavors of restoration of the social contract by more fiscal support. Nevertheless, in 2014 and under the pressure of the World Bank to lend Egypt $15 Billion to recover from the economic stagnation phase, Egypt initiated the first increase in the gasoline and diesel prices in the MENA by 60%. Not only Egypt but also Qatar, UAE, Kuwait, and Saudi Arabia followed suit. Oil prices have hurt the finances of MENA oil-exporting countries, requiring them to initiate or intensify the reform of energy subsidies. Consequently, the price rise was a must for these countries to keep up with their current prosperity, and this had been partially achieved by reducing the fiscal breakeven oil price between 2014 and 2016. [2, 3]. A greater focus on decoupling economic growth from energy consumption will have a long-lasting impact on fiscal revenues for net-oil exporting countries. Moreover, the consequences of the decoupling will allow more renewables penetration in the current energy mix enabling many countries to reach their Paris Agreement goals. For short-term energy policy actions, starting a subsidy reform towards the final repeal of subsidies is a must as these measures relate to all end-use sectors and impact fiscal stability in many countries.

## 9. RESULTS – RENEWABLE TARGETS

Egypt has the most ambitious PV targets in the region, mainly due to accommodating population growth and expected industrial developments. In 2014, Egypt produced only 9% from renewables, mainly due to the High Dam in Aswan "Hydropower". Its 2020 goal is to reach 20% with 14% primary energy consumption with 220 MW from PV. In 2019, the Egyptian government commissioned a 1.65GW PV Benban Solar Park. That made it the only country in the region to reach its 2020 target. For example, Jordan failed to meet its 2020 targets twice, with an expected 7% in 2015 and 10% in 2020. renewable generation. That was way below the current status, with only 1% from renewable resources. A complete list of the targets and their status is summarized in Table 3.

## 10. CONCLUSION

Although the MENA region will continue to be a significant oil and gas producer, consumer, and exporter for years to come, a global transition to new energy supply sources is underway, which will continue to impact all MENA countries. Specifically, the increasing adoption of alternative energy internationally and a greater focus on decoupling economic growth from energy consumption will have a lasting impact on export revenues for net-oil exporting countries. Rapidly increasing regional energy demand coupled with increasingly limited low-cost oil and gas supply in some MENA countries, and the global shift towards low-carbon energy, necessitate an energy system transition across the region.

Such policies impact countries' abilities to reach their 2020 and 2030 renewable targets. Egypt has started a subsidy reform shifting its spending from subsidizing fossil fuel products to developing and commission new renewable projects. Benban Solar Park in Aswan is an example of such a shift, allowing its energy mix to incorporate 1.65GW from PV. On the other hand, Jordan could not perform this shift due to political instability causing failure to reach 2015 and 2020 targets.

Near-term energy policy actions should include subsidy reform and progress toward the ultimate elimination of subsidies as these measures relate to all end-use sectors and impact fiscal stability in many countries. Although renewable energy is a central topic of energy system diversification globally, there is a significant interest in nuclear energy and coal to complement the greater use of natural gas in the power sector. As a final policy consideration, regional energy cooperation is essential but must be approached thoughtfully and realistically. However, electricity market integration in the GCC may be viable within the next several years if GCC countries maintain their focus on least-cost energy system development and make further strides in electricity price normalization. As the GCC electricity market makes progress, further connections to North Africa may eventually be possible. However, a truly integrated electricity market across the MENA is hard to foresee soon.

## 12. APPENDIX

| Country | Expected Population Growth in Millions | | | Expected Electricity consumption per capita (kWh/capita) | | |
|---|---|---|---|---|---|---|
| | 2015 | 2020 | 2030 | 2014 | 2020 | 2030 |
| Egypt | 93,78 | 102,94 | 119,75 | 1658,00 | 2088.66 | 2647,00 |
| Jordan | 9,16 | 10,21 | 11,12 | 1888,00 | 2304,00 | 2779,62 |
| UAE | 9,15 | 9,81 | 11,06 | 11264,00 | 10117,00 | 8897,00 |
| Saudi Arabia | 31,56 | 34,71 | 39,48 | 9444,00 | 10443,00 | 12936,00 |
| Bahrain | 1,37 | 1,70 | 2,01 | 19592,00 | 19358,22 | 19358,22 |
| Kuwait | 3,94 | 4,30 | 4,87 | 15213,00 | 16030,22 | 16030,22 |
| Oman | 4,20 | 5,15 | 5,90 | 6554,00 | 8926,00 | 10589,90 |
| Qatar | 2,48 | 2,79 | 3,23 | 15309,00 | 15162,18 | 15162,18 |
| Iran | 79,36 | 83,59 | 88,86 | 2986,00 | 3576,40 | 4589,60 |
| Iraq | 36,12 | 41,50 | 53,30 | 1306,00 | 1157,95 | 1206,18 |
| Algeria | 39,87 | 43,33 | 48,82 | 1356 | 1530,00 | 1983,00 |
| Libya | 6,23 | 6,66 | 7,34 | 1857,00 | 2652,00 | 2652,00 |
| Morocco | 34.80 | 37.07 | 40.87 | 901,00 | 1091,00 | 1392,00 |
| Tunisia | 11,27 | 11,90 | 12,84 | 1444,00 | 1655,00 | 2012,00 |
| Lebanon | 5,85 | 6,02 | 5,37 | 2893,00 | 2985,26307 | 2985,26307 |

**Table 1.** Values of the expected Electricity consumption per capita according to the population growth

| Country | Status | Visions | Announced Target | | | | Announced Projects |
|---|---|---|---|---|---|---|---|
| | 2014 | | 2020 | | 2030 | | |
| | Share of Renewable Energy | | Renewable Energy | PV | Renewable Energy | PV | |
| Egypt | 9% | Egypt Vision 2030 | 14% Primary Energy 20% Electricity generation | in 2020: 220 MW | 13.5 GW | N/A | Al Wadi Al Gedid - 6MW - in bid stage - Masdar Komombo - 200 MW - EOI Phase - NREA FIT Round 1 |
| Jordan | 0.36% | Jordan Vision 2025 | 7% in 2015 (2007-2020 - Failed) 10% in 2020 (2007-2020 - Failed) 11% in 2025 Final Energy Production | 600MW (Failed) 300 MW | N/A | N/A | Renewables Energy and Energy Efficiency Law (REEL) Shams Ma'an solar PV farm 100 MW - 300M US$ Quweira PV Power Plant - 65 |
| | | | | | | | MW - PQ Stage<br>MEMR Round 3 - 200 MW |
| UAE | 0.27% | Vision 2021 | National Target - 1.6 GW<br>Abu Dhabi 7% Electricity Generation<br>Dubai 7% Electricity Generation | N/A | 44% in 2050<br>Dubai 25% Electricity Generation - 1 GW | 2.5 GW Rooftops PV (20%) | 1000 MW Noor Abu Dhabi 1 Solar PV farm<br>13 MW Dubai Solar PV Project in operation - First stage for the 1000 MW project<br>ADWEA Sweihan - 350 MW - |
| Saudi Arabia | 0% | Vision 2030 | 4% | 3.45 GW (unspecified type) | 50% - 54 GW in 2032 | 16 GW | Duba Phase I Integrated solar combined cycle - 50 MW - PQ Stage |
| Bahrain | 0% | N/A | 5% | N/A | N/A | N/A | N/A |
| Kuwait | 0% | Kuwait 2035 | 5% in 2015 Electricity Generation (Failed)<br>10% in 2020 | N/A | 15% | 4.6 GW | Shagaya Multi-Technology Renewable Energy Park 2GW<br>10 MW Solar PV was tendered in 2013 |
| Oman | 0% | Vision 2020 | N/A | N/A | N/A | N/A | Royal Oman Police Hospital - 1 MW - Bid Preparation |
| Qatar | 0% | Vision 2030 | 2% | 1.8 GW in 2015 (Failed)<br>1.8 GW in 2018 - 20% | 20% | | a tender was lunched in 2014 for 1.8 GW<br>Al Aymadi solar project - 1 MW - Bid Preparation |
| Iran | 5.2% | N/A | N/A | N/A | N/A | N/A | N/A |
| Iraq | 4:3% | N/A | 2% in 2016 (Failed) | 240 MW in 2016 (Failed) | N/A | 10% | N/A |
| Algeria | 0.4% | Vision 2030 | 6% in 2015 (Failed)<br>15% in 2020 | N/A | 40% Final Energy<br>27% Electricity Generation | 37% (PV & CSP)<br>13.5 GW | N/A |
| Libya | 0% | N/A | 3% in 2015 (Failed)<br>10% Primary Energy<br>7% Electricity Generation<br>10% in 2025 Electricity Generation | 129 MW in 2015 (Failed)<br>344 MW in 2020<br>844 MW in 2025 | N/A | N/A | Hoon PV Power Plant - 14 MW - in bid stage |
| Morocco | 12.3% | N/A | 2000 MW - 42% | 2 GW | 52% | N/A | 2,000MW of solar capacity across five sites at an anticipated cost of $9 billion Noor IV PV - 80 MW - Bid Preparation |
| Tunisia | 3.09% | N/A | 11% in 2016 Electricity Generation (Failed)<br>16% in 2016 Power Capacity (Failed) | N/A | 25% Electricity Generation<br>40% Power Capacity | 10 GW | N/A |
| Lebanon | 1.08% | N/A | 12% | N/A | N/A | N/A | W |

**Table 2.** Announced Governmental targets and projects
Sources: Clean Energy Pipeline, Griffiths 2017, EY MENA Solar Outlook

| Country | Policies in action | | Missing policies | | Financing Credit Rating | | |
|---|---|---|---|---|---|---|---|
| | Regulatory measure | Fiscal Incentives | Regulatory measure | Fiscal Incentives | S&P | Moody's | Fitch |
| Egypt | Feed in Tarif (Under Revision)<br>Net Metering<br>Tendering | Capital Subsidy or grant<br>Reduction in Sales (VAT) | Tradable REC<br>Electric utility quota | production tax credits<br>Public investment or loans | CCC+ | Caa1 | B- |
| Jordan | Feed in Tarif (0,163US$/kWh for PV)<br>Net Metering<br>Tendering | Reduction in Sales<br>Public investment loans | Tradable REC<br>Electric utility quota | Capital subsidy or grant<br>production tax credits | BB- | B1 | N/A |
| UAE | Electricity Utility Quota<br>Net Metering<br>Tendering | Energy Production Payment<br>Public | Tradable REC<br>Feed in Tarif | Capital subsidy or grant<br>Production tax credits | AA | Aa2 | AA |
| | | Investment Loans | | Reductions in sales | | | |
| Saudi Arabia | N/A | N/A | N/A | N/A | AA- | Aa3 | AA- |
| Bahrain | N/A | Public Investment Loans | N/A | Capital subsidy or grant production tax credits | BBB | Baa1 | BBB- |
| Kuwait | Tendering | N/A | Tradable REC<br>Feed in Tarif<br>Electric Utility Quota<br>Net Metering | N/A | AA | Aa2 | AA |
| Oman | N/A | N/A | N/A | N/A | A | A1 | N/A |
| Qatar | N/A | N/A | N/A | N/A | AA | Aa2 | N/A |
| Iran | Feed in tariff | Investment Tax Credit<br>Energy Production Payment<br>Public Investment | Tradable REC<br>Electric Utility Quota<br>Net Metering | Reductions in sales | N/A | N/A | B+ |
| Iraq | Tendering | N/A | Tradable REC<br>Feed in Tarif<br>Electric Utility Quota<br>Net Metering | N/A | N/A | N/A | N/A |
| Algeria | Feed in tariff<br>Tendering | Capital Subsidy or grant<br>Energy Production Payment<br>Public Investment Loans | Tradable REC<br>Electric Utility Quota<br>Net Metering | Production tax credits<br>Reduction in Sales | N/A | N/A | N/A |
| Libya | N/A | Reduction in sales | N/A | Capital subsidy or grant<br>Production tax credits | N/A | N/A | B |
| Morocco | Net Metering<br>Tendering | Public Investment Loans | Tradable REC<br>Feed in Tarif<br>Quota Obligation | Capital subsidy or grant<br>Production tax credits | BBB- | Ba1 | BBB- |
| Tunisia | Net Metering | Capital subsidy or grant<br>Reduction in Sales<br>Public investment loans | Tradable REC<br>Feed in Tarif<br>Quota Obligation | production tax credits | BB- | Ba2 | BB+ |
| Lebenon | Net Metering | Reduction in Sales<br>Public investment loans | Tradable REC<br>Feed in Tarif<br>Quota Obligation | Energy Production Payment<br>Production tax Credits<br>Capital subsidy | B | B1 | B |

**Table 3**. Current renewable Energy Policies policies and Financing Credit
Sources: Clean Energy Pipeline 2017, Griffiths 2017